\documentclass[12pt]{article}
\usepackage{latexsym}
\usepackage{amsmath}
\usepackage[dvips]{graphicx}
\usepackage{textcomp}

\begin{document}

\title{Neutron Imaging by Boric Acid}
\author{Fabio Cardone$^{1,2}$, Giovanni Cherubini$^{3}$,\\ Walter Perconti$%
^{6}$,Andrea Petrucci$^{2,4,\ast}$, Alberto Rosada $^{5}$\\
\\
$^{1}$Istituto per lo Studio dei Materiali Nanostrutturati (ISMN --- CNR) \\
Via dei Taurini - 00185 Roma, Italy \\
$^{2}$GNFM, Istituto Nazionale di Alta Matematica ''F.Severi'\ \\
Citt\`a Universitaria, P.le A.Moro 2 - 00185 Roma, Italy \\
$^{3}$ARPA Radiation Laboratories \\Via Montezebio, 01100 Viterbo, Italy \\
$^{4}$Italian National agency for new technologies, Energy and\\
sustainable economic development ( ENEA ),\\ Via Anguillarese, 301 -
00123 Roma, Italy
 \\
$^{5}$ Unit\`a Tecnica Tecnologie e Impianti \\per la Fissione e la
Gestione del Materiale Nucleare(UTFISST),\\ Laboratorio
Caratterizzazione Materiali Nucleari(CATNUC), \\Italian National
agency for new technologies, Energy and \\ sustainable economic
development (ENEA),\\ Via Anguillarese, 301 - 00123 Roma, Italy
\\
$^{6}$Istituto Superiore per la Protezione e la Ricerca Ambientale
(ISPRA),\\ Via Vitaliano Brancati, 48 - 00144 Roma, Italy\\
 $^{\ast}$ Corresponding author: petrucciandr@gmail.com}
\date{}
\maketitle

\abstract{In this paper a new type of passive neutron detector based
on the already existing one, CR39, is described. Its operation was
verified by three different neutron sources: an Americium-Beryllium
(Am241-Be) source; a TRIGA type nuclear reactor; and a fast neutron
reactor called TAPIRO. The obtained results, reported here,
positively confirm its operation and the accountability of the new
developed detecting technique.}

\section{Introduction}
The methods used to detect neutrons are commonly based on the
emission of charged particles produced by the interaction of
neutrons with He-3 or BF$_3$ contained in the counters or detectors.
Another type of commonly used detectors is the BD, the Defender and
the DefenderXL which are Bubble Detector-Dosimeters. They are made
of an elastic polymer gel in which minute droplets of superheated
liquid are uniformly scattered. When these droplets are struck by
neutrons, they evaporate and form small bubbles of gas that remain
trapped in the elastic polymer. The number of bubbles is in direct
proportionality with the number and the energy of neutrons, in
particular from the counted number of bubbles in these detectors and
from their certified efficiency (Bubbles/mSv) one can obtain the
dose equivalent of the neutrons that struck the detector. A further
technique is based on CR39. The CR39 (PADC) are plastic detectors
made of the polymer C$_{12}$H$_{18}$O$_7$, whose density is 1.3 g
cm$^{-3}$. They can detect both heavy charged particles and neutrons
when the latter are converted into alpha particles by the nuclear
reaction B-10 ( n,$\alpha$ ) Li-7 that takes place within a very
thin homogeneous layer of B-10 that coats the surface of the
CR39~\cite{tommasino}. These detectors have a very wide range of
energy extending from tens of KeV (40 KeV) up to some MeV (4MeV). A
further technique to measure neutrons is by activation of targets
made of thin plates of ultra-pure Au, In, Co, Cu, Mn, Ni, Al, Ti
followed  by the measurement of the gamma and beta activity of these
plates. However, this technique is possible only with quite high
fluence of neutrons. The detecting process described in this paper
is based on a technique used by the authors in previous experiments
in which polycarbonate plates (CR39) were surrounded by a layer of
suitable thickness of granular boric acid
H$_3$BO$_3$~\cite{piezoneutr}. In the present case, the
polycarbonate plates were substituted for a normal photographic film
made of an acetate substrate with a layer of silver halide spread on
it. This film was surrounded by a layer of H$_3$BO$_3$ of suitable
thickness as it was for the plates. In both cases, the grains of
boric acid have the same dimension. Due to a patent pending on this
technique we are not allowed to provide further details. This method
is based on the same technique described above where neutrons are
converted by the boric acid into lithium 7 and alpha particles by
the reaction

\begin{equation}
^{10}B^5_5  + n \rightarrow ^7Li^4_3 + \alpha + 2.31 MeV\footnote{The available energy is devided between the alpha particle and the Lithium atom.
 When the latter is left in a non-excited state (about the 6\% of the time) the available energy is 2.792 MeV
 while when the Lithium atom is left in an excited state (about the 94\% of the time) the available energy is 2.31
 MeV.}
\end{equation}

 and the latter interact with a polycarbonate plate on which they
produce peculiar tracks.

\section{Description and preparation of the detectors}
\begin{figure}
\begin{center} \
\includegraphics[width=0.8\textwidth]{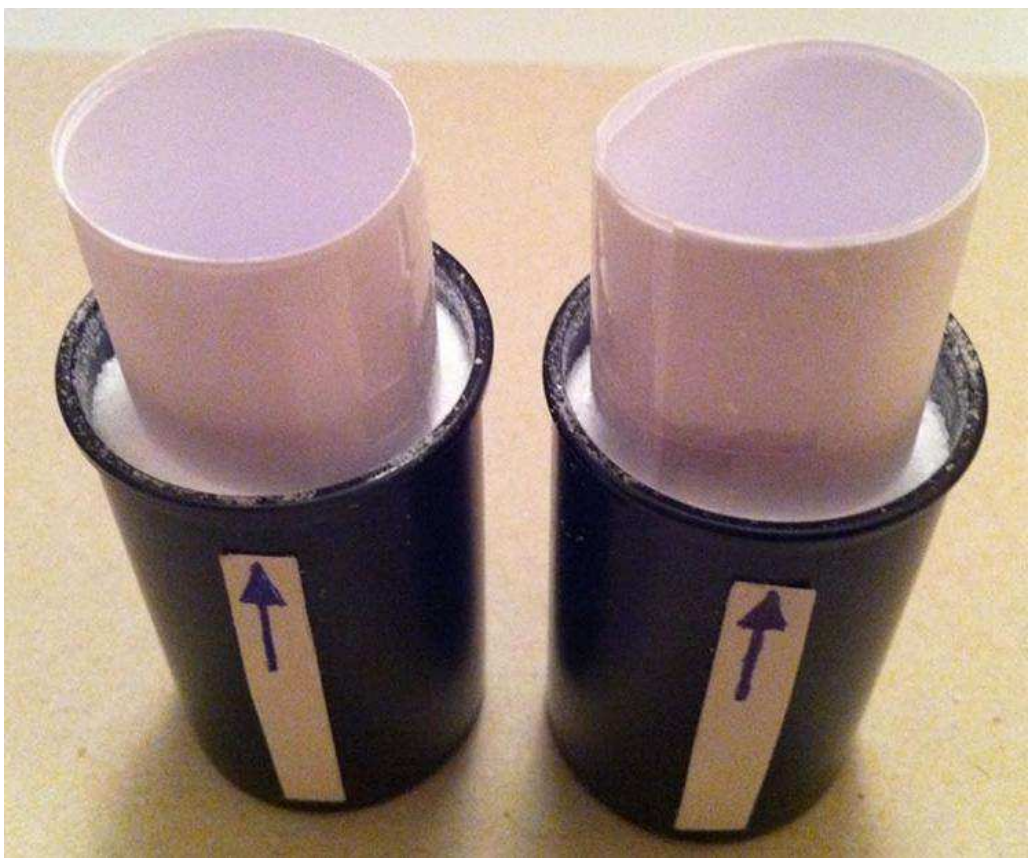} \caption{The black cylinder contains a cylindrical layer of H$_3$BO$_3$
(the white granulate) which is in direct contact with the film
(lucid external part of the white cylinder in this figure). More
H$_3$BO$_3$ will be poured into the cylindrical remaining space and
once that all the black cylinder is full of H$_3$BO$_3$, the
cylindrical sheet of paper (visible in this figure), which only
gives more rigidity to the film, is slipped away. }\label{film_box}
\end{center}
\end{figure}

According to the reaction mentioned above, we developed a specific
type of detector in which a usual black and white photographic film
of 400 ISO made of a layer of silver halide was surrounded by a
layer of boric acid. The choice to use a 400 ISO film was made in
order to have a compromise between the brightness and the resolution
of the image. The detector was made with few frames of photographic
film, 24x36 mm, wound in order to form a cylinder by joining the two
edges. The second step was to insert this cylindrical film inside a
dark cylinder of Polyethylene and then fill this cylinder of boric
acid as shown in Fig.\ref{film_box}. All of the different tasks to
set up the detector were made inside a dark room in complete
obscurity, that is without turning on the inactinic light.

\section{Experimental Measurements}
We carried out measurements with three films of the same type
described above for each test. In each measurement the three parts
of the film were subjected to different conditions: the first part
was kept in air and was not irradiated; the second part was kept in
boric acid and it was not irradiated; the third part, was immersed
in boric acid and was put in front of the neutron flux. These
experiments were carried out at the ENEA Casaccia Research Centre,
with three types of neutron sources:
Am241-Be source; horizontal channel in the thermal column of the
nuclear reactor TRIGA; radial channel 2 of the nuclear reactor
TAPIRO.

\begin{table}[tbp]
\caption{Main general characteristics of the available neutron
sources. $\Phi$ is the flux of the corresponding neutron source}
\label{neutronsources}
\begin{center}
\begin{tabular}{|ll|ll|ll|ll|}
\hline \multicolumn{2}{|c|}{---} &
\multicolumn{2}{|c|}{\textbf{Source Am-241-Be}}&
\multicolumn{2}{|c|}{\textbf{TRIGA RC1}} &
\multicolumn{2}{|c|}{\textbf{TAPIRO}}
\\ \hline
\multicolumn{2}{|c|}{$\Phi$ neutron} & \multicolumn{2}{|c|}{389 n
cm$^{-2}$ sec$^{-1}$} & \multicolumn{2}{|c|}{---} &
\multicolumn{2}{|c|}{---}
\\\hline
\multicolumn{2}{|c|}{$\Phi$ Therm horiz channel} &
\multicolumn{2}{|c|}{---} & \multicolumn{2}{|c|}{194 n cm$^{-2}$
sec$^{-1}$} & \multicolumn{2}{|c|}{---}

\\\hline
\multicolumn{2}{|c|}{$\Phi$ Radial channel 2} &
\multicolumn{2}{|c|}{---} & \multicolumn{2}{|c|}{---} &
\multicolumn{2}{|c|}{278 n cm$^{-2}$ sec$^{-1}$}
\\\hline
\multicolumn{2}{|c|}{R$_{Cd}$} & \multicolumn{2}{|c|}{---} &
\multicolumn{2}{|c|}{2.2} & \multicolumn{2}{|c|}{---}
\\ \hline
\multicolumn{2}{|c|}{fixed dose rate ($\mu$Sv/h)} &
\multicolumn{2}{|c|}{100} & \multicolumn{2}{|c|}{100} &
\multicolumn{2}{|c|}{100} \\ \hline \multicolumn{2}{|c|}{available
channels} & \multicolumn{2}{|c|}{---} & \multicolumn{2}{|c|}{12} &
\multicolumn{2}{|c|}{7}
\\ \hline
\multicolumn{2}{|c|}{maximum power} & \multicolumn{2}{|c|}{---} &
\multicolumn{2}{|c|}{1 MW thermic} & \multicolumn{2}{|c|}{5 kW
thermic}
\\ \hline
\multicolumn{2}{|c|}{cooling system} & \multicolumn{2}{|c|}{---} &
\multicolumn{2}{|c|}{Water convection} &
\multicolumn{2}{|c|}{Helium}
\\ \hline
\multicolumn{2}{|c|}{nuclear fuel} & \multicolumn{2}{|c|}{---} &
\multicolumn{2}{|c|}{Enriched U 20$\%$} &
\multicolumn{2}{|c|}{Enriched U 93.5$\%$}
\\ \hline

\end{tabular}
\end{center}
\end{table}

\begin{figure}
\begin{center} \
\includegraphics[width=0.8\textwidth]{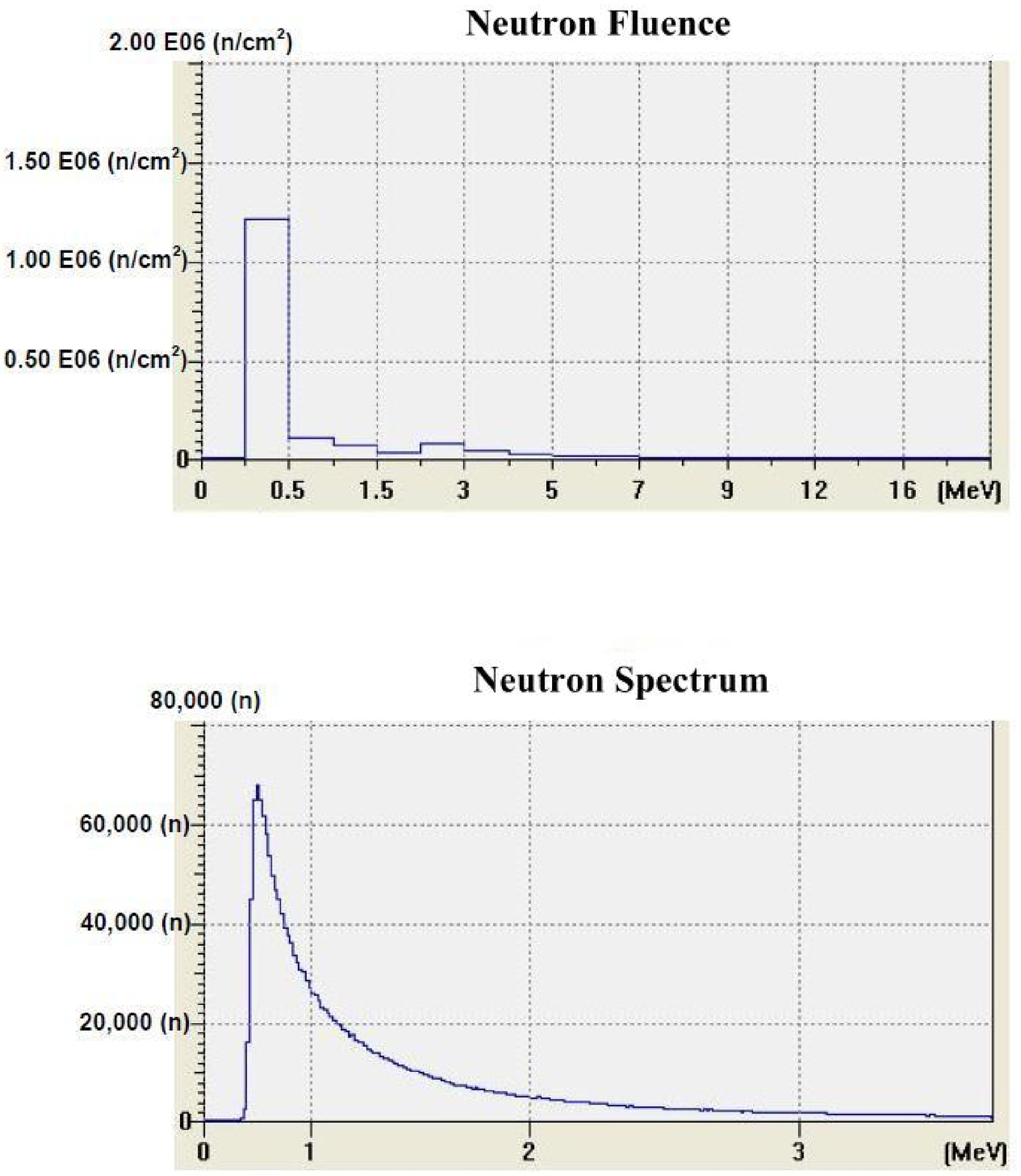} \caption{Fluence and neutron spectrum of the Am-241 - Be source:
100 $\mu$Sv/h. Duration of the measurement: 1 hour. Counts per
second: 400 cps. Measuring instrument: Neutron Spectrum MicroSpec-2
Neutron Probe (Bubble Technology Industry).}\label{AmBespectr}
\end{center}
\end{figure}

\begin{figure}
\begin{center} \
\includegraphics[width=0.8\textwidth]{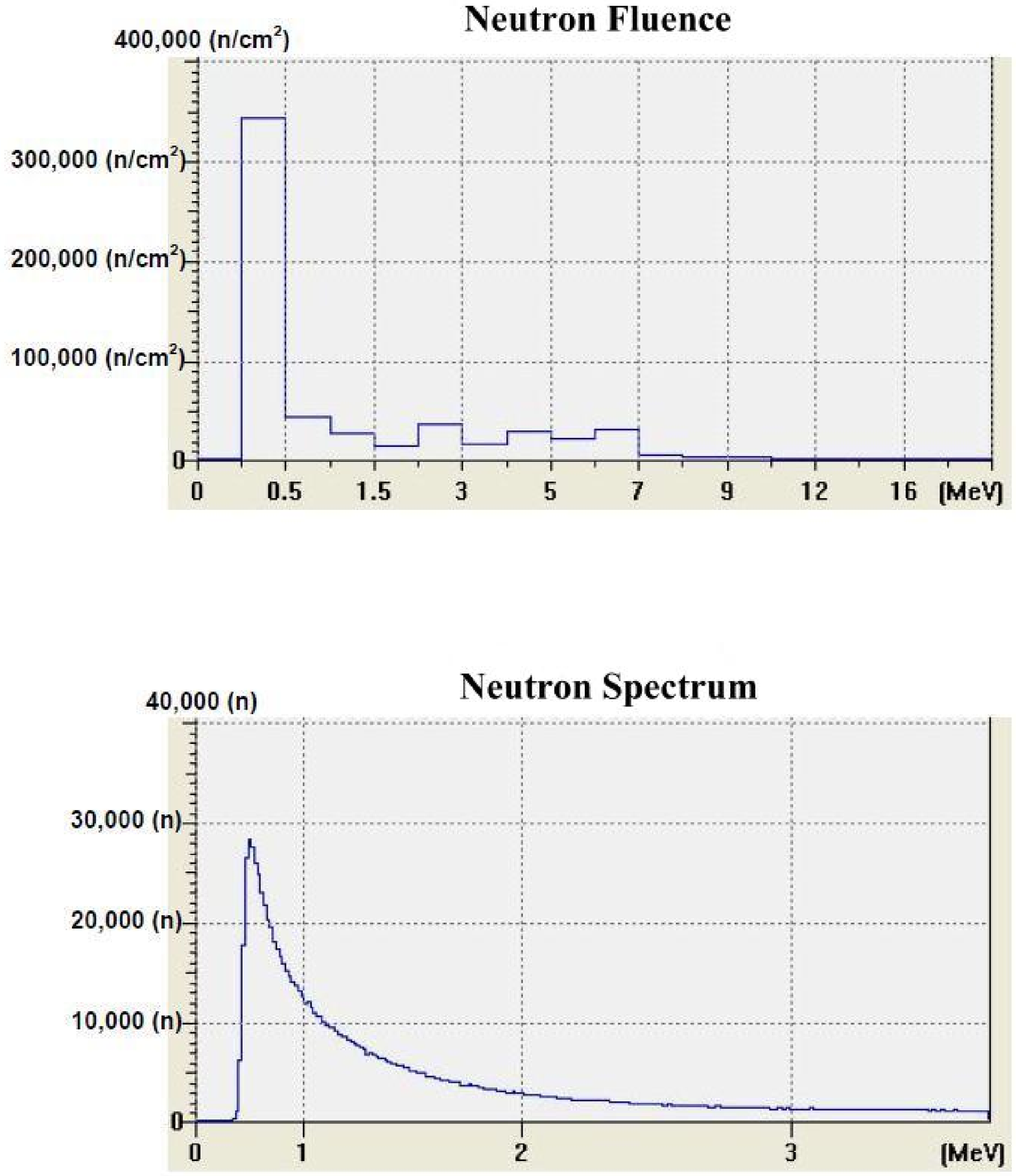} \caption{Fluence and neutron spectrum of the source TRIGA:
thermal column horizontal channel. Dose Rate: 100 $\mu$Sv/h.
Duration of the Measurement: 1 hour. Counts per second: 1000 cps.
Measuring instrument: Neutron Spectrum MicroSpec-2 Neutron Probe
(Bubble Technology Industry).}\label{TRIGAspectr}
\end{center}
\end{figure}

\begin{figure}
\begin{center} \
\includegraphics[width=0.8\textwidth]{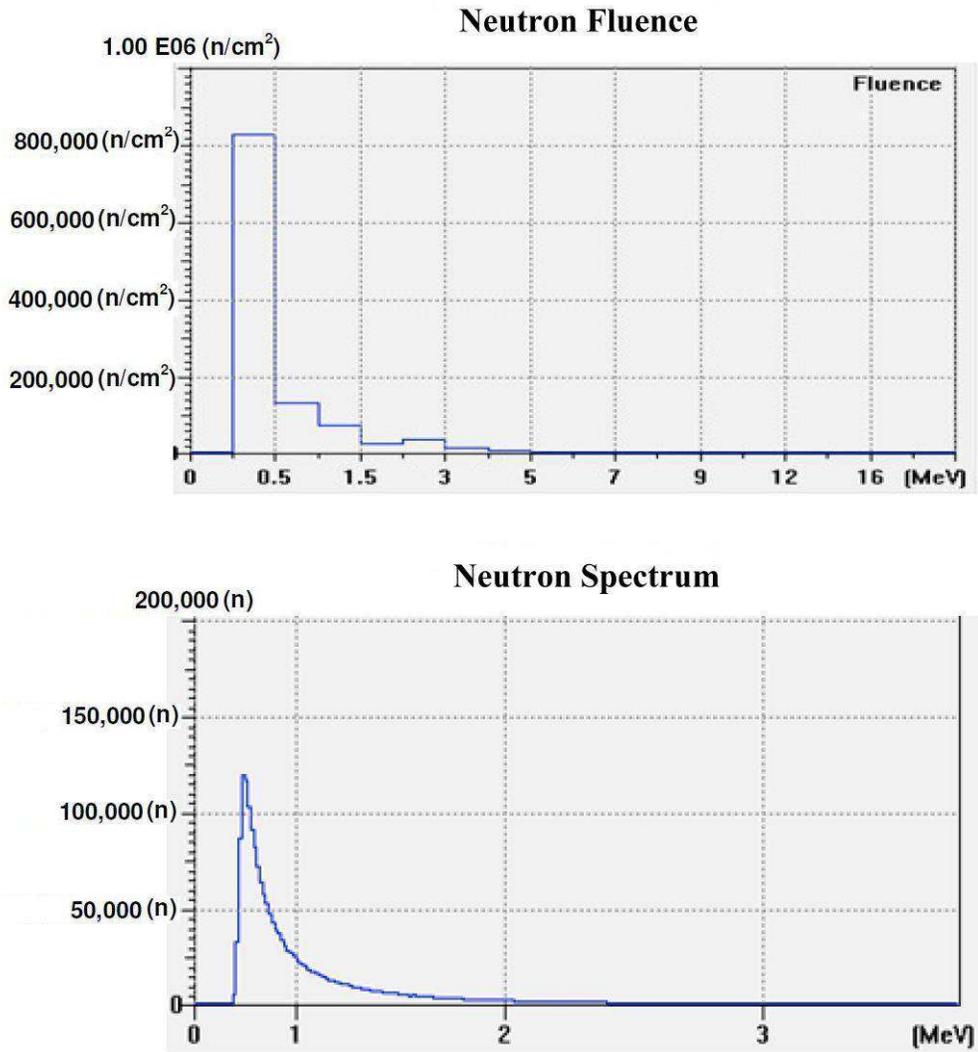} \caption{Fluence and neutron spectrum of the source TAPIRO:
radial channel 2. Dose Rate: 100 $\mu$Sv/h. Duration of the
Measurement: 1 hour. Counts per second: 30 cps. Measuring
instrument: Neutron Spectrum MicroSpec-2 Neutron Probe (Bubble
Technology Industry).}\label{TAPIROspectr}
\end{center}
\end{figure}

The irradiation time was 1 hour (3600 s) and the dose rate of
neutrons was 100 $\mu$Sv/h for all of the irradiated film. The dose
rate had been previously ascertained by an He-3 dosimeter which was
placed in the same position where the boric acid detectors will have
been placed. In Fig.\ref{AmBespectr} we report the characterisation
of the Am-241 - Be source as to the fluence of its neutrons and its
neutron spectrums. The characterisation was carried out by the
neutron spectrometer MicroSpec-2~\cite{microspec2} Neutron
Probe~\cite{neutrprobe} (Bubble Technology Industry) which comprises
two detectors ( He-3 and liquid scintillator NE213 ) for two neutron
energy intervals. In Figs.\ref{TRIGAspectr} and \ref{TAPIROspectr}
we report the fluence and the spectrum, which were measured by the
same spectrometer, of the neutrons from the reactors TRIGA and
TAPIRO. In all the three measurements, the films with H$_3$BO$_3$,
after development, presented visible and clear traces which,
conversely, were present neither on the control film screened by
H$_3$BO$_3$, nor on the control film in air that were not
irradiated. By comparing this experience with a previous one in
which we irradiated, in similar conditions, CR39 plates with
neutrons in front of the radial channel 2 of the reactor TAPIRO, we
find out a very peculiar similarity between the traces obtained on
the CR39 screened by boric acid and those obtained on the silver
halide of the film screened by boric acid. This detecting technique
produces different images for different neutron sources, but for the
same neutron source, the images are the same independently of the
type of detector used (CR39 or film with silver halide). At the end
of the irradiations, the boric acid that had played the role of
converting material (it converted neutron into alpha particles) for
the irradiated detectors or of a simple screen for the non
irradiated ones, was analysed by mass spectrometry. This analysis
was conducted in order to look for possible traces of the
neutron-into-alpha converting reactions (see 1). In these reactions,
the isotope of Boron with higher capture cross section for thermal
and slightly epithermal neutrons is B-10. When it captures a
neutron, it splits yielding an alpha particle and a nucleus of
Lithium-7. Thus, by mass spectrometry we searched both for a
possible change in the natural occurring ratio of Boron-11 and
Boron-10, expecting a slight decrease of the lighter isotope and a
corresponding increase of Lithium-7.

\section{Results}
\begin{figure}
\begin{center} \
\includegraphics[width=0.4\textwidth]{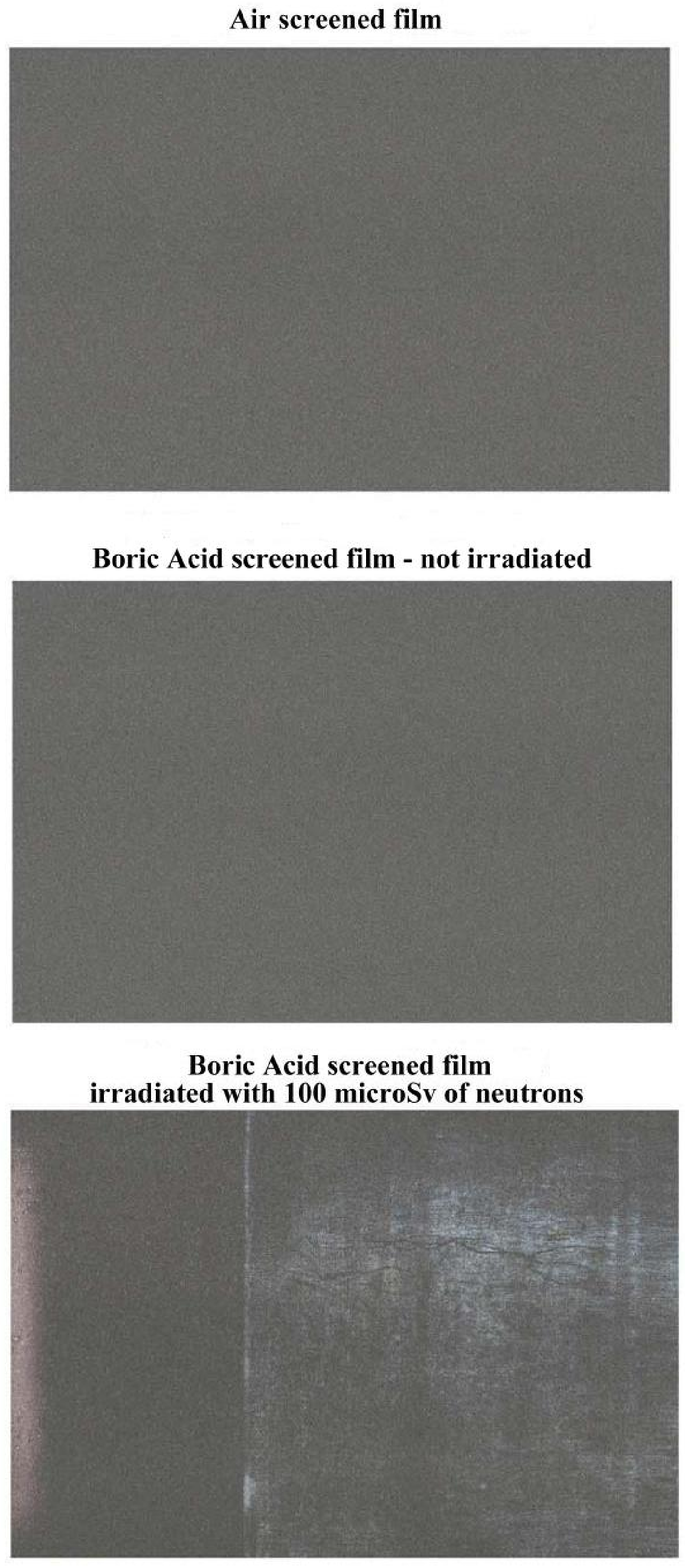} \caption{Neutron Detection from the Am-241 - Be source.
Silver halide black and white film: sensitivity: 400 ISO; granular
H$_3$BO$_3$}\label{AmBe_neutr_photo}
\end{center}
\end{figure}

\begin{figure}
\begin{center} \
\includegraphics[width=0.4\textwidth]{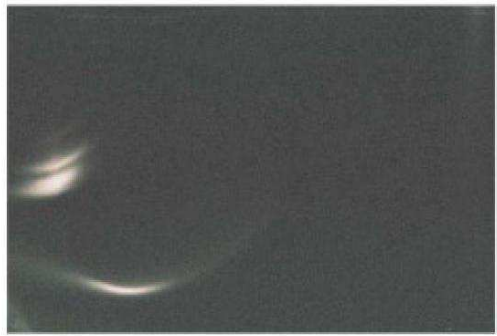} \caption{Neutron Detection from the reactor TAPIRO: radial channel 2.
Silver halide black and white film: sensitivity: 400 ISO; granular
H$_3$BO$_3$}\label{TAPIRO_neutr_photo}
\end{center}
\end{figure}

\begin{figure}
\begin{center} \
\includegraphics[width=0.5\textwidth]{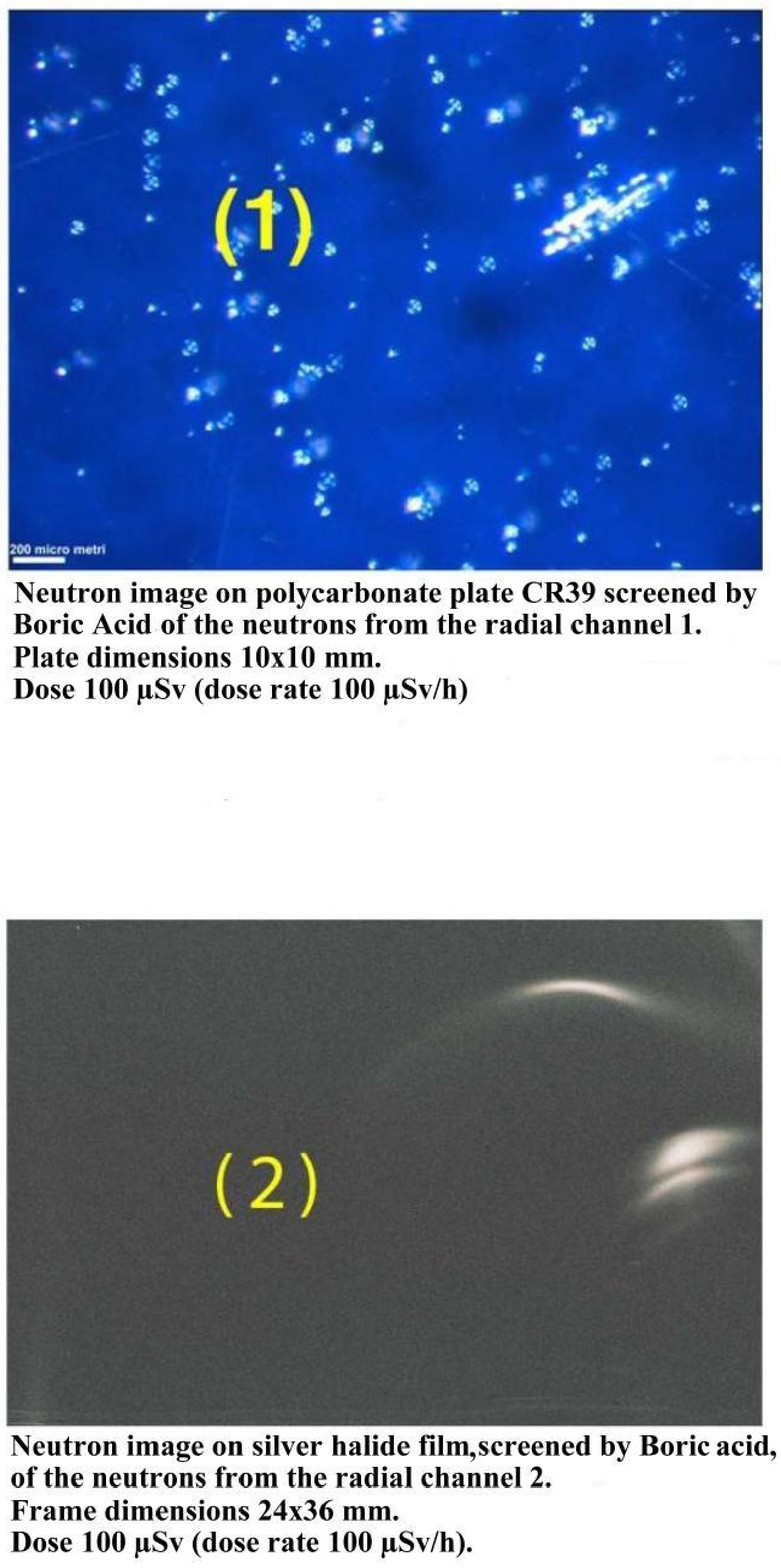} \caption{Comparison between two neutron images obtained on two different
imaging substrates screened by the same filter by neutrons of the
reactor TAPIRO from radial channel 1 and radial channel 2. The two
images have the same morphology.}\label{film_cr39}
\end{center}
\end{figure}

Each of the Figs.\ref{AmBe_neutr_photo}, \ref{TRIGA_neutr_photo} and
\ref{TAPIRO_neutr_photo} comprises three images: the first is the
non irradiated film in air; the second is the non irradiated film
immersed in boric acid; the third is the irradiated film screened by
boric acid. In Fig.\ref{film_cr39}, there is the visual comparison
between two images: the tracks on the CR39 and those on the acetate
film with silver halide. These images were obtained by the same
source (reactor TAPIRO) and with the same thickness of boric acid
surrounding the CR39 and the acetate film. Despite the different
substrates (polycarbonate and silver halide) the morphology of the
tracks is the same. With regards to the analyses by mass
spectrometer, we found out a fairly significant variation of the
ratio between B-11 and B-10 in the detectors irradiated by neutrons.
In particular, we found out the two following ratios between
Boron-11 and Boron-10 for the non-irradiated Boron and for the
irradiated one, respectively.\\
\begin{center}
$\frac{^{11}B}{^{10}B}$= 3.9906 $\pm$ 4$\cdot$10$^{-4}$
$\rightarrow$ non irradiated\\
\end{center}

\begin{center}
$\frac{^{11}B}{^{10}B}$= 3.9981 $\pm$ 4$\cdot$10$^{-4}$
$\rightarrow$ irradiated\\
\end{center}

\begin{center}
($\frac{^{11}B}{^{10}B}$)$_{irradiated}$ $>$
($\frac{^{11}B}{^{10}B}$)$_{non-irradiated}$\\
\end{center}

Despite this piece of evidence, nothing is possible to say about the
presence of Lithium since, even with a significant variation of
B-10, the corresponding variation of Lithium-7 was below the minimum
value that could be discerned by the mass spectrometer.

\section{Conclusion}
\begin{figure}
\begin{center} \
\includegraphics[width=0.4\textwidth]{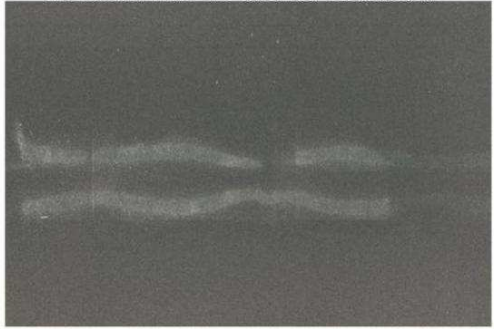} \caption{Neutron Detection from the reactor TRIGA: thermal column horizontal channel.
Silver halide black and white film: sensitivity: 400 ISO; granular
H$_3$BO$_3$}\label{TRIGA_neutr_photo}
\end{center}
\end{figure}

We developed a new and cheap technique which is capable to detect
neutron emission and we verified its accountability by testing it
with different sources. These tests produced different images for
different sources possessing the same dose rate
(Figs.\ref{AmBe_neutr_photo}, \ref{TRIGA_neutr_photo} and
\ref{TAPIRO_neutr_photo}). Besides, the same neutron source with
equal screening thickness of boric acid surrounding the revealing
substrate, produces similar images on different substrates
(Fig.\ref{film_cr39}) independently of the revealing substrate
(polycarbonate and silver halide). A  further proof of the
accountability of this technique is available in
Fig.\ref{TRIGA_neutr_photo} where the third picture shows the shape
of the neutron collimating horizontal slit at the end of the thermal
column channel.

\section{Aknowledgements}
We want to thank for their kind, precious and active collaboration
the following people: Giovanni Silvestri owner of the Hobby Photo
Laboratories, based in Sulmona (Italy) for the preparation of the
film and its development; Massimo Sepielli managing director of the
UTFISST ENEA; the team of the nuclear reactor TRIGA of the ENEA
Casaccia Research Centre, Daniele Baiano (BAS IONIRP ENEA), Monica
Lammardo (UTFISST-REANUC ENEA)reactor operator, Emilio Santoro
(UTFISST-REANUC ENEA) director of the reactor TRIGA ; the team of
the nuclear reactor TAPIRO of the ENEA Casaccia Research Centre
Barbara Bianchi (UTFISST-REANUC ENEA), Orlando Fiorani
(UTFISST-REANUC ENEA) and Alfonso Santagata (UTFISST-REANUC ENEA).

\end{document}